\documentstyle[twocolumn,aps,epsfig,graphicx,epsf]{revtex}

\draft

\begin{document}
\title{High-harmonic generation from a confined atom}
\author{Carla Figueira de Morisson Faria and Jan-Michael Rost}
\address{Max-Planck-Institut f\"{u}r Physik komplexer Systeme,\\
 N\"{o}thnitzer Str. 38, 01187 Dresden, Germany}
\date{\today}
\maketitle

\begin{abstract}
The order of high harmonics emitted by an atom in an intense laser field
is limited by the so-called cutoff frequency. Solving the time-dependent
Schr\"odinger equation, we show that this frequency
can be increased considerably by a parabolic confining  potential, if the
confinement parameters are suitably chosen.
 Furthermore, due to confinement, the radiation intensity
remains high throughout the extended emission range. 
All features observed can be explained with classical arguments.
\end{abstract}



Typical features of the emission spectra of an atom in a strong laser field,
known as ``the plateau'' and ``the cutoff'', are a wide frequency region
with harmonics of comparable intensities, and an abrupt intensity decrease
at the high-energy-end of the plateau. For a monochromatic driving field,
the cutoff energy is given by $\varepsilon _{\max }=|\varepsilon
_{0}|+3.17U_{p}$, where $|\varepsilon _{0}|$ and $U_{p}$ are the ionization
potential and the ponderomotive energy, respectively. This simple cutoff
law, derived by classical means only \cite{Cor93,tstclass}, or using more
refined methods \cite{tstep}, corresponds to the physical picture referred
to as the ``three-step model'' \cite{Cor93,tstclass,tstep}: A bound electron
exposed to the laser field leaves the atom through tunneling at a time $%
t_{0} $ (step 1), propagates in the continuum, being driven back towards its
parent ion at a later time $t_{1}$ (step 2), and finally falls back to a
bound state under emission of high harmonics (step 3). This scenario
describes the spectral features observed experimentally very well \cite{expe}%
. The cutoff frequency, in quantitative agreement with the experiment, is
related to the maximum kinetic energy the electron has upon return, $E_{{\rm %
kin}}(t_{1},t_{0})$.

According to this picture, in order to increase the cutoff energy, one must
increase the kinetic energy of the returning electron. Indeed, the exisiting
proposals to extend the plateau towards higher energies reach a higher value
of $E_{{\rm kin}}(t_{1},t_{0})$ by different means. However, this does not
necessarily imply an efficient generation of high-order harmonics up to this
larger cutoff energy.

For instance, a rather complex situation with several ``cutoffs '' \cite{cf4}
emerges by using bichromatic fields with driving waves of comparable
intensities. An illustrative example is presented in \cite{cf5}, using a
driving field of linearly polarized monochromatic light of frequency $\omega 
$ and its second harmonic. Under such conditions the monochromatic cutoff,
as a function of the field-strength ratio between the two driving waves,
splits into two branches. Thereby, the upper branch extends up to $%
|\varepsilon _{0}|+5U_{p}$. However, the harmonics emerging up to the cutoff
of the upper branch are weak compared to those from the lower branch and
therefore irrelevant to the emission spectrum. The reason is simple: The
intensity of the harmonics is strongly influenced by step 1 which is the
tunneling process out of the binding potential under the influence of the
field. If the field amplitude is small at the emission time $t_{0}$ (which
is the case for the upper branch) then the tunneling barrier is large and
the generated harmonics will be weak compared to those which originate from
an effective tunneling process (as it is the case for the lower branch).

Another idea to increase the cutoff energy is to use a static electric
field. It provides an additional force which accelerates the electron
towards the atomic core resulting in a higher kinetic energy $E_{{\rm kin}%
}(t_{1},t_{0})$. Indeed, it has been demonstrated that with an electric
field whose strength is only a few percent of the amplitude of the laser
field one can considerably enlarge the cutoff energy \cite{static,magn}.
However, the scheme suffers from two principal limitations. First, the
increased kinetic energy occurs mainly for electrons with long excursion
times. Due to wave packet spreading, those trajectories have negligible
influence on the harmonic spectra. This problem has been overcome by
introducing an additional magnetic field to restrict the spreading \cite
{magn}. A second, more severe limitation is the pronounced bound-state
depletion caused by the static electric field: the atom is irreversibly 
ionized within a few field cycles, such that no appreciable high-harmonic
generation takes place.

The bound-state depletion which prevents an effective extension of the
high-harmonic frequency points to the principle dilemma easily described in
the picture of the returning electron: To extend the plateau and increase
the cutoff, a kinetic energy of the returning electron, as large as
possible, is desirable. On the other hand, an electron with such a high
energy will leave the atom and is lost for the possible generation of high
harmonics in consecutive laser cycles.

Hence, we need a mechanism which brings an electron back to the nucleus,
despite the fact that it has a kinetic energy so high that it would be
irreversibly driven away from the core. Naively, a simple wall for the
electron should already do this. However, one must avoid that the abrupt
reflection of the charged electron at a wall leads to Bremsstrahlung which
masks the desired high-harmonic generation of the atom.

In the following we will show that the idea of bringing back the fast
electron by an additional confinement and thereby extending the cutoff for
the spectrum without additional depletion does indeed work for a suitably
soft confinement potential.

We consider a one-dimensional situation, which is a reasonable approximation
for linearly polarized light. Atomic units are used throughout. The binding
of the electron is described by the potential 
\begin{equation}
V_{{\rm a}}\left( x\right) =-1.1\ \exp \left( -x^{2}/1.21\right) ,
\end{equation}
which supports a single bound state $|0\rangle $ at energy $\varepsilon
_{0}=-0.58\ {\rm a.u}$., corresponding to the Argon ionization potential.
The system is exposed to a monochromatic laser field $E(t)=E_{0}\sin \omega t
$ and the additional confining potential (Fig.1)
\begin{eqnarray}
V_{{\rm h}}(x) &=&\frac{\Omega _{{\rm h}}^{2}}{2}x^{2}h(x),  \eqnum{2a}
\label{trappot} \\
h(x) &=&\left\{ 
\begin{array}{c}
1,|x|<x_{0} \\ 
\cos \left( \frac{\pi }{2}\theta \right) ,\ x_{0}\leq |x|\leq x_{\max } \\ 
\ 0,|x|>x_{\max }
\end{array}
\right.   \eqnum{2b}
\end{eqnarray}
with $\theta =(|x|-x_{0})/(x_{\max }-x_{0})$. The parameter $%
x_{0}=nE_{0}/\omega ^{2}$ is chosen to be a multiple of the electron
excursion amplitude, and $x_{\max }=2x_{0}.$  Parabolic potential shapes are
taken as a first approximation in several physical systems, as for instance
electromagnetic traps \cite{trap} or solid-state devices \cite{solid}. 
Note, that for the parameter
range chosen, identical emission spectra are obtained with and without
truncation of the harmonic potential indicating that even in the truncated
potential depletion has negligible influence. Thus, the electron does not
reach the edges of $V_{{\rm h}}\left( x\right) $, which indicates an
effective confinement$.$ Futhermore, this shows that the confining potential
does not generate harmonics itself. Therefore, high-harmonic generation
still takes place only near the atomic core, for which the coordinate $x$ is
considerably smaller than the electron excursion amplitude.
\begin{figure}[tbp]
\vspace{-0.2cm}
\noindent
\epsfxsize  7.9cm
\epsffile{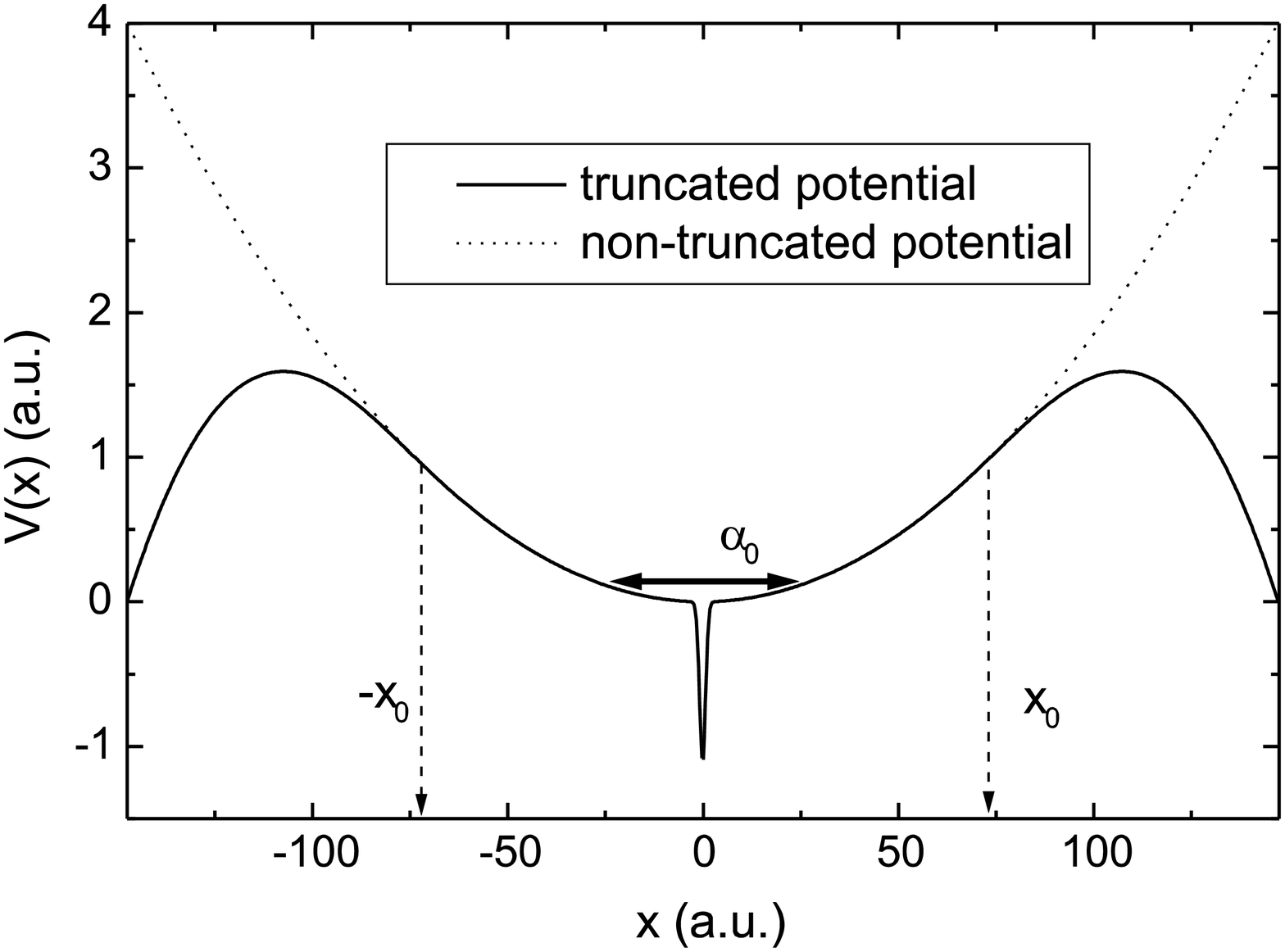}
\caption{Schematic representation of an atom in an external confining potential 
$V_{{\rm h}}(x)$ (c.f. Eq. (\protect{\ref{trappot}})). The parameter $x_0$ for which the potential is 
truncated and the electron excursion amplitude $\alpha_0$, for the parameters
of Fig.2(a), as
well as the non-truncated
potential, are indicated in the figure.}
\end{figure}

The evolution of the electronic wave packet is described by the
time-dependent Schr\"{o}dinger equation 
\begin{equation}
i{\frac{d}{dt}}|\psi (t)\rangle =\left[ {\frac{p\sp{2}}{2}}+V(x)-p\cdot
A(t)\right] |\psi (t)\rangle ,
\end{equation}
with $V(x)=V_{{\rm a}}\left( x\right) +V_{{\rm h}}\left( x\right) ,$ and the
emission spectra are given by 
\begin{equation}
\label{sptdse}
\sigma (\omega )=\left| \int_{0}^{\infty }d(t)\exp [-i\omega t]\right| ^{2},
\end{equation}
where the dipole acceleration 
\begin{equation}
d(t)=\left\langle \psi (t)\right| -dV(x)/dx+E(t)\left| \psi (t)\right\rangle 
\end{equation}
is computed by means of Ehrenfest's theorem \cite{dipacc}. We take the atom
initially in the ground state $|0\rangle $.
 \begin{figure}[tbp]
\noindent
\epsfxsize  7.9cm
\epsffile{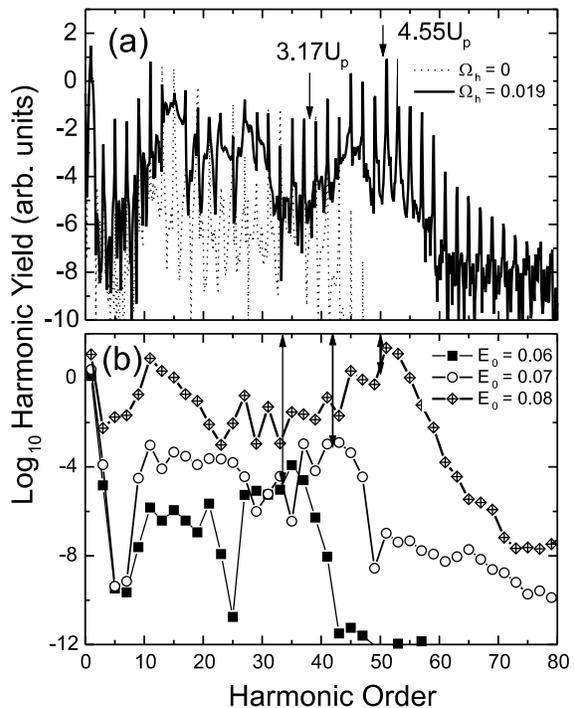}\\
\caption{ Harmonic spectra calculated using the TDSE
(c.f. Eq. (\protect{\ref{sptdse}})). Part (a): Field amplitude $%
E_{0}=0.08\ {\rm a.u.},$ field frequency $\omega =0.057\ {\rm a.u}.,$ without
(dashed line) and with (solid line) confinement ($\Omega _{{\rm h}}=0.019\ {\rm a.u}). $ Part (b): Field strengths  $E_{0}=0.06\ {\rm a.u}.,\
E_{0}=0.07\ {\rm a.u.}$ and $E_{0}=0.08\ {\rm a.u.,}$ confinement curvature  $\Omega _{{\rm h}}=0.019\ {\rm %
a.u}.$ and the same frequency as in the previous part.  The classical cutoff
energies, given by the cutoff law $|\varepsilon _{0}|+4.55U_{p}$, correspond
to the harmonic orders $n =33,\ n =41$ and $n =49$
 and are indicated by arrows in the figure. In part (b), only the
harmonic intensities are given, connected by lines.}
\end{figure} Furthermore, in the results to
be presented we chose $x_{0}=73.87\ {\rm a.u.},$ which corresponds to three
times the excursion amplitude of an electron in a monochromatic field with $%
E_{0}=0.08\ {\rm a.u.}$ and $\omega =0.057\ {\rm a.u.}$.
 With these field
parameters and a reasonable choice of $\Omega _{{\rm h}},$ one indeed finds
that the high-harmonic spectrum extends beyond the cutoff energy $%
\varepsilon _{\max }=|\varepsilon _{0}|+3.17U_{p}$ without significant loss
of intensity, see Fig.~2.
More specifically, we have determined a cutoff
energy of $|\varepsilon _{0}|+4.55U_{p}$ which is a 50\% increase compared
to the case without trapping. The classical argument for the cutoff energy
applies to the situation with confinement as well and we find very good
agreement between the cutoff in the quantum spectra (e.g. Fig.~2) and the
classical cutoff. The latter has been determined in analogy to the situation
without confinement: Starting with an electron of velocity zero, its
trajectory is propagated under the influence of the laser field and the
confinement potential $V_{{\rm h}}$ (but without the atomic potential $%
V(x)=V_{{\rm a}}$). We vary the initial time $t_{0}$ for which the electron
leaves the atom within a field cycle, computing $E_{{\rm kin}}(t_{1},t_{0})$
for return times $t=t_{1}$ satisfying the condition $x(t_{1})=0.$ The local
maxima in $E_{{\rm kin}}(t_{1},t_{0})$ yield the classical prediction for
the cutoffs in the harmonic spectra.

The good agreement of the classical cutoff with the one found in the quantum
spectra allows us to predict, with the classical model, the behavior of the
cutoff as a function of the external parameters, i.e. the confinement
constant $\Omega _{{\rm h}}$, the frequency and the amplitude of the
external field. We find that in the parameter range of interest the cutoff
law can be written in the form 
\begin{equation}
\varepsilon _{\max }=|\varepsilon _{0}|+f(\Omega _{{\rm h}},\omega )U_{p},
\label{concut}
\end{equation}
where $f(\Omega _{{\rm h}},\omega )$ in general neither exhibits a simple
functional form nor can be derived analytically. However, the linear
dependence on the field intensity $E_{0}^{2}$ through $U_{p}$ in Eq.~(\ref
{concut}) is preserved just as in the case without confinement, see Fig.~3.
Only for large confinement constants or electron trajectories with long
excursion times 
 $f(\Omega _{{\rm h}},\omega )$ becomes slightly intensity-dependent.
\vspace{-0.1cm}
\begin{figure}[tbp]
\noindent
\epsfxsize  7.5cm
\epsffile{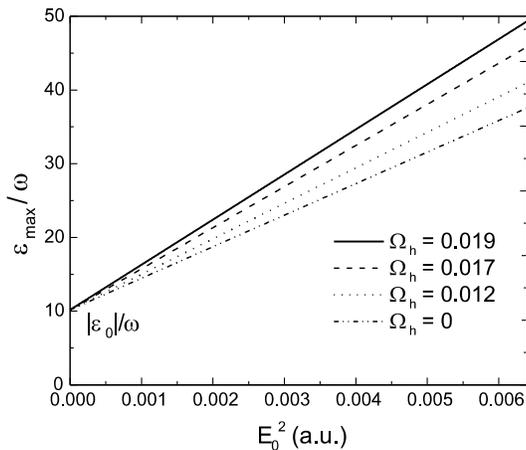}
\caption{Cutoff energies computed using the classical model, as functions of
the field intensity $E_0^2 $, for confinement constants $0\leq
\Omega_{{\rm h}} \leq 0.019\ {\rm a.u..}$ and 
$\omega =0.057\ {\rm a.u.}$. }
\end{figure}
The general behavior of $f(\Omega _{{\rm h}},\omega )$ is rather complex.
Nevertheless, asymptotically a simple and familiar behavior is recovered:
For very high frequency, the monochromatic cutoff constant is is approached,
i.e., $f(\Omega _{{\rm h}},\omega \to \infty )\to 3.17$, as can be seen in
Fig.~4. For finite frequency $\omega $ the cutoff energy increases with
growing $\Omega _{{\rm h}}$. In fact, the lower the frequency, the more
sensitively the cutoff law depends on $\Omega _{{\rm h}}$. This property is
the actual reason why one can obtain an increased cutoff energy with a
confinement. For very low frequencies, the cutoff energy can be easily
extended beyond $|\varepsilon _{0}|+9U_{p}.$ In practice, however, there is
a lower frequency limit to generate an appreciable intensity of high
harmonics in the present context. If the confinement frequency is comparable
to the laser frequency, $\Omega _{{\rm h}}\sim \omega $, the confinement
potential itself starts to contribute to the harmonic generation process,
ceasing to be a passive element. Hence, the condition for HHG under a
confinement potential can be written as $\Omega _{{\rm h}}/\omega \ll 1$.
However, there is also the usual upper limit in frequency $\omega $ which
comes from the requirement that the atom in the laser field must be in the
tunneling regime \cite{footnote1}.
\vspace{-0.1cm}
\begin{figure}[tbp]
\noindent
\epsfxsize  7.5cm
\epsffile{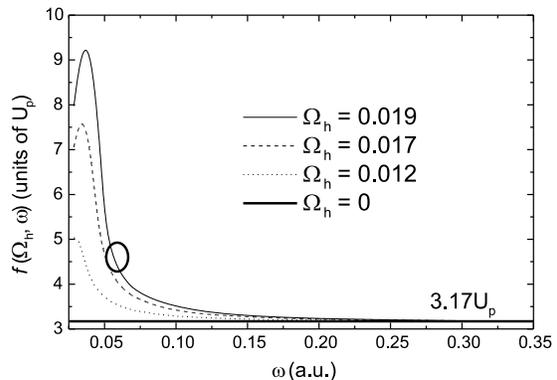}
\caption{Cutoff energies computed using the classical model, as functions of
the frequency $\omega $ of the driving field, for confinement curvatures $0\leq
\Omega_{{\rm h}} \leq 0.019\ {\rm a.u..}$ The circle in the figure corresponds to $%
\omega =0.057\ {\rm a.u.}$ and $\Omega _{{\rm h}}=0.019\ {\rm a.u.,}$ for
which the spectra in Fig.~2 have been calculated. }
\end{figure}

Typical frequencies used in HHG experiments, and for which a long plateau is
obtained, are in the vicinity of $\omega =0.057\ {\rm a.u.}$. For this
frequency a confinement indeed leads to a larger cutoff energy as
demonstrated in Fig.~2.

In conclusion, we have presented a new scheme for increasing the cutoff
energy of the high-harmonic spectra of an atom under the influence of a
strong laser field. Placing the atom in a confining parabolic potential, we
have shown that the cutoff energy can be increased by more than fifty
percent. An effective increase of the cutoff requires a careful choice of
the confinement strength. The confinement curvature $\Omega _{{\rm h}}$ must
be strong enough for the electron to be appreciably accelerated towards the
parent ion, but weak enough for it to move in a ``quasi-continuum''. If  $%
\Omega _{{\rm h}}$ is too weak, the conventional cutoff law $|\varepsilon
_{0}|+3.17U_{p}$ is not altered by it. If $\Omega _{{\rm h}}$ is too strong,
the electron moves as a bound particle that does not generate higher
harmonics. In the extreme case, one observes the dipole response of a
harmonic oscillator, i.e., equally spaced resonances. A rough indication of
whether the electron is in a ``quasi-continuum'' is given by the ratio of
the energy difference between two consecutive levels of the confinement
potential, $\Delta \varepsilon _{{\rm h}}=\Omega _{{\rm h}},$ and the
ionization potential of the atom in question. If $\Omega _{{\rm h}%
}/|\varepsilon _{0}|\ll 1,$ this condition is fulfilled. Also, as already
discussed, the ratio between the frequency $\omega $ of the external field
and the confinement curvature $\Omega _{{\rm h}}$ plays an important role.
If $\Omega _{{\rm h}}/\omega \sim 1,$ the parabolic potential contributes
too actively to the harmonic generation process, and the plateau and cutoff
are not present in the spectra. The best results have been obtained for $%
x_{0}\sim 100\ {\rm a.u}.,$ $\Omega _{{\rm h}}\sim 0.02\ {\rm a.u}.$ and $%
\omega \sim 0.04\ {\rm a.u}.$ In this case, the energy difference between
two consecutive levels of the confinement potential is still of the order of
one tenth of the ionization potential $|\varepsilon _{0}|$ and $\Omega _{%
{\rm h}}/\omega \sim 0.5$. For this parameter range,
 the cutoff energy can be extended until approximately $|\varepsilon_{0}|+6U_p$.

On a more technical level, yet very interesting from the theoretical point
of view, we have seen that the cutoff law is given by the classical picture
of an electron moving under the influence of the laser field and the
confinement potential. Very good agreement between the quantum-mechanical
full calculation and the classical model occurs for a wide range of field
strengths, frequencies around $\omega \sim 0.05\ {\rm a.u}.$ and confinement
curvatures of the order of $\Omega _{{\rm h}}\sim 10^{-2}\ {\rm a.u}.$
Thereby we have found that the cutoff law strongly depends on the
confinement curvature $\Omega _{{\rm h}}$ and the frequency $\omega $ of the
laser field, but only linearly on the field intensity $E_{0}^{2}$.

The proposed setup presents several advantages over the schemes using static
fields. For instance, using a confining potential, one can achieve a
considerable extension of the cutoff energy already for the trajectories
corresponding to {\it short} electron excursion times, whereas using static
fields one mainly affects electron trajectories with {\it long} excursion
times. Due to wave-packet spreading, the former trajectories are far more
important for the harmonic spectra than the latter. In order to reduce the
spreading one needs very strong magnetic fields \cite{magn}. Another
noteworthy feature of a confinement potential is that one can obtain
stronger harmonics than in the static field, or even in the monochromatic
case. In fact, a serious disadvantage concerning static electric fields is
an appreciable decrease in the harmonic intensities compared to the field
free case, due to depletion, i.e. irreversible ionization. This problem is
not present in our scheme.

However, similarly to the so far proposed extension of the cut off energy
 by using a combination
of a static electric and magnetic fields, we are not
aware of a direct possibility for an experimental realisation of our scheme.
 In the former case the magnetic
field necessary is unrealistically large for a laboratory application 
\cite{magn2}. For
our situation, a true electromagnetic trap is too macroscopic compared to
the paramater range we need. On the other hand there might be exciting
possibilities in the future to design a confined atom as described in a
quantum-dot like device, for instance as an impurity. An important issue
here, however, is the limitation in the radiation intensity in order to
avoid the damage threshold. Recently, solid-state materials
which can survive our parameter range, namely fields of
wavelength $\lambda =790\ {\rm nm}$ and intensities above
 $10^{14}\ {\rm W/cm^2}$, have been observed \cite{damage}.

Acknowledgements: We would like to thank K. Richter, D. B. Milo\v {s}evi\'{c},
M. L. Du and K. Leo for useful discussions.



\begin{references}
\bibitem{footnote1}  The so-called ``tunneling regime'' is characterized by
a Keldysh parameter $\gamma =\sqrt{|\varepsilon _{0}|/2U_{p}}$ smaller than
one.
\end{references}
\end{document}